\documentclass{jpp}
\usepackage{graphicx}
\usepackage{amsmath}
\usepackage{amssymb}
\usepackage{bm}
\usepackage{color}
\usepackage{physics}
\usepackage{natbib}
\usepackage{comment}
\usepackage[normalem]{ulem}

\usepackage{makecell}

\newcommand{\beq}{\begin{equation}}
\newcommand{\eeq}{\end{equation}}

\newcommand{\benum}{\begin{enumerate}}
\newcommand{\eenum}{\end{enumerate}}

\newcommand{\abra}[1]{\left\langle{#1}\right\rangle} 
\def\bar#1{\overline{#1}}
\newcommand{\del}{{\bm{\nabla}}}

\newcommand{\bmB}{{\bm{B}}}
\newcommand{\bmb}{{\bm{b}}}
\def\bmu{\bm{u}}
\newcommand{\bmU}{{\bm{U}}}
\newcommand{\bmk}{{\bm{k}}}

\newcommand{\emf}{{\mathcal{E}}}

\newcommand{\nuM}{{\nu_\text{M}}}

\newcommand{\bmcalU}{\bm{\mathcal{U}}}
\newcommand{\alphaK}{{\alpha^\text{K}}} 
\newcommand{\anglebracket}[1]{{\langle{#1}\rangle}}

\newcommand{\ks}{k_\text{S}}

\shorttitle{Anisotropic quenching}
\shortauthor{H. Zhou and E. G. Blackman}

\title{Generalized quenching of large-scale magnetic dynamos in anisotropic flows}

\author{
Hongzhe Zhou\aff{1,2}
	\corresp{\email{hzhou21@ur.rochester.edu}},
Eric G. Blackman\aff{1,2}
	\corresp{blackman@pas.rochester.edu}}

\affiliation{\aff{1}Department of Physics and Astronomy, University of Rochester, Rochester, NY, 14627, USA
\aff{2}Laboratory for Laser Energetics, University of Rochester, Rochester NY, 14623, USA}

\begin{document}

\maketitle

\begin{abstract}
The buildup of small-scale magnetic helicity which accompanies the oppositely signed growth on large scales is central to conventional dynamical quenching theories of mean-field dynamos. 
{However, the conventional formalism presumes isotropy and thereby excludes part of the magnetic Lorentz back-reaction for anisotropic turbulence, rendering it insufficient to predict the full quenching for such flows.}
To overcome this deficiency, we derive a new generalized quenching formalism that includes the full back-reacting Lorentz force
even for anisotropic flows, and a new ``selective-damping-$\tau$'' closure which conserves magnetic helicity.
We apply the formalism to examples of $\bm\alpha^2$ dynamos and show its predicted quenching for different cases of turbulence---isotropic helical, anisotropic helical, and anisotropic non-helical. It predicts stronger-than-conventional quenching in general, but reduces to the conventional case in the helical isotropic limit.
\end{abstract}

\section{Introduction}
Mean field dynamo (MFD) theory is commonly used to explain amplification and sustenance of large-scale magnetic fields in stars \citep{RaedlerWiedemannBrandenburg1990,KukerRudiger1999,ChabrierKuker2006, BlackmanThomas2015,BlackmanOwen2016}, 
galaxies \citep{RSS1988, Shukurov2006galacticFountain, SSS2007, Chamandy2013a, Chamandy2013b, Rodrigues2019}, 
and accretion disks \citep{RuedigerElstner1995, VishniacBrandenburg1997, Rekowski2000, Blackman2012}.
It focuses on the dynamics on the ``large scales'' which are comparable to the size of the system. Correlations of ``small-scale'' fluctuations (usually turbulent) are coarse-grained and modeled statistically.

The widely invoked ``$\alpha$ effect'' is a prominent example of an MFD mechanism,
for which assuming isotropy makes $\alpha$ a pseudoscalar that measures how much turbulent electromotive force (EMF) is aligned with the mean magnetic field.
In the simplest case, $\alpha$ is proportional to the difference between the small-scale kinetic helicity and current helicity \citep{Pouquet1976,BlackmanField2002prl,BrandenburgSubramanian2005}, i.e.,
\beq
\alpha\propto\bar{\bmu\cdot\del\times\bmu}-\bar{\bmb\cdot\del\times\bmb},
\label{eqn:alpha_iso}
\eeq
where $\bmu$ and $\bmb$ are respectively the turbulent velocity and magnetic fields, and the overline indicates an ensemble average.
Astrophysical rotators plausibly have ample kinetic helicity to drive a MFD because their large-scale density gradient and axis of rotation are non-orthogonal.
In galaxies the two are aligned, and in stellar convection zones the density gradient points to the center of the star whereas the axis of rotation has a uniform direction.
In such turbulent flows, rising blobs expand and falling blobs contract as they conserve angular momentum, maintaining a constant sign of kinetic helicity in each hemisphere.
In this case, the growth rate of the large-scale magnetic energy is proportional to the (time-dependent) magnitude of $\alpha$, and the dynamo saturates when $\alpha$ depletes to a value that it just balances other destructive process present, such as turbulent diffusion, flow advection, or magnetic reconnection.


The current helicity term in equation (\ref{eqn:alpha_iso}) is responsible for $\alpha$ depletion, and thus dynamo saturation or ``quenching.'' 
The reason why $\bmb$ must have a component aligned with its curl is due to the conservation of magnetic helicity which is defined as the volume integral of the dot product between the magnetic field and its vector potential.
The small-scale current helicity is proportional to the small-scale magnetic helicity in a Coulomb gauge in a two-scale approach for closed volume, and has to grow with an opposite sign to that of the growing large-scale magnetic helicity.
 As the small scale current helicity grows in turn, it offsets the kinetic term in kinetically forced systems \citep{Pouquet1976,BlackmanField2002prl, BrandenburgSubramanian2005, BlackmanSubramanian2013}.
We refer to this understanding of quenching as ``conventional quenching'' in this paper.
It is presently widely applied in practical models \citep{Shukurov2006galacticFountain, SSS2007}.

The conventional quenching formalism invokes isotropy of the transport coefficients, 
but realistic turbulent flows are likely anisotropic due to finite scale separation, vectors coherent on large scales such as rotation or density stratification, or gradients in local sources of kinetic helicity \citep{RadlerKleeorinRogachevskii2003, RadlerStepanov2006, Brandenburg2013}.
For example, kinetic helicity depends on the angle between the density gradient and the rotation axis as in stellar convection zones \citep{ZhouBlackman2019}, and even in the case of galaxies its strength can depend on the local rate of supernova explosions which is the turbulent energy supply \citep{ZhouBlackman2017}.

Once we relax the assumption of isotropy, 
the ``$\bm\alpha$''-effect (notice the switch to bold font) then refers to a rank-2 pseudo-tensorial coefficient in mean-field equations which is composed of correlation functions of $\bmu$ and $\bmb$, and only its trace is directly related to the kinetic and current helicities.
In the isotropic case, $\bm\alpha$ is completely determined by its trace, but for an anisotropic flow, $\bm\alpha$ obtains off-diagonal elements which are generally independent of its trace (although we will see that for incompressible flows the spectrum of $\bm\alpha$ has only one pseudo-scalar degree of freedom).
A generalized quenching theory to allow for anisotropic flows is therefore needed, even to assess the extent to which lessons learned from the standard isotropic case apply. 


{The $\bm\alpha$ effect may operate in cohort with other dynamo mechanisms in realistic systems (e.g., the $\Omega$ effect, shear-current effect, etc.), but as we will show and emphasize, anisotropic $\bm\alpha$ quenching must include the entire Lorentz force back-reaction, a principle that stands independent of any of these other mechanisms that may also be operating. 
Toward this end in the present paper, we isolate the $\bm\alpha$ effect and}
answer the following question for an $\bm\alpha^2$ dynamo:
what is the the quenching mechanism in anisotropic multi-scale turbulent flows, for which helicity conservation alone is insufficient to dynamically connect the evolution of 
$\bm\alpha$ to growing mean-fields?
In section \ref{sec:2} we develop such a general quenching formalism using a ``selective-damping-$\tau$'' closure that conserves magnetic helicity.
In section \ref{sec:3} we show that anisotropic and even non-helical (i.e. traceless) $\bm\alpha$ tensors also allow for exponential growth of the mean magnetic field, and show the incompleteness
of conventional quenching in this case and how our new 
 quenching formulae predict stronger quenching when applied to such a dynamo.
We conclude in section \ref{sec:conclusion}.

\section{Time evolution of $\bm \alpha$ from the Full Lorentz Backreaction}
\label{sec:2}
\newcommand{\tot}[1]{{#1}^\text{tot}}
\subsection{Expression for $\bm\alpha$ in mean-field theory}

The mean-field formalism of electrodynamics can be found in e.g., Refs. \cite{Moffatt1978} and \cite{BrandenburgSubramanian2005}. In this subsection we use a particular formalism to derive $\bm\alpha$ that facilitates bridging the gap between previous and our present work.
For incompressible flow, the Navier-Stokes and induction equations for the velocity field $\bmU$ and magnetic field $\bmB$ are, respectively,
\beq
\partial_t\bmU
=\hat{\bm P}\left(
-\bmU\cdot\del\bmU
+\bmB\cdot\del\bmB
\right)
+\nu\nabla^2\bmU
+{\bm f}^\text{K},
\label{eqn:NS}
\eeq
and
\beq
\partial_t\bmB
=\del\times\left(
\bmU\times\bmB
\right)
+\nuM\nabla^2\bmB
+{\bm f}^\text{M},
\label{eqn:induction0}
\eeq
where magnetic fields are measured in Alfv\'en units, 
$\hat P_{ij}=\delta_{ij}-\partial_i\partial_j/\nabla^2$ is the projection operator which eliminates the pressure term, 
$\nu$ is the kinematic viscosity, $\nuM$ is the magnetic resistivity,
and ${\bm f}^\text{K}$ and ${\bm f}^\text{M}$ are respectively the kinetic and magnetic forcing functions.
When driving occurs only through the Navier-Stokes equation as in the cases of stellar convection zones and galaxies, 
${\bm f}^\text{M}$ vanishes, but it becomes important in magnetically driven dynamos, e.g., those in coronae of stars and accretion disks.
Here we assume periodic boundaries in all three directions, and use indices $(1,2,3)$ for $(x,y,z)$.

We decompose velocity and magnetic fields into a large-scale mean and a small-scale fluctuation, i.e.,
\newcommand{\K}[1]{{#1}^\text{K}}
\newcommand{\M}[1]{{#1}^\text{M}}
\newcommand{\C}[1]{{#1}^\text{C}}
\beq
\bmU=\bar\bmU+\bmu,\ 
\bmB=\bar\bmB+\bmb.
\eeq
Reynolds rules are satisfied when large scale separation is assumed
\citep{BrandenburgSubramanian2005,ZBC2018JPP} 
so that 
\beq
\bar\bmu=\bm0,\ 
\bar{\bar\bmU}=\bar\bmU,\ 
\bar{\bar\bmU\bmu}=\bar\bmU\ \bar\bmu=\bm0,
\eeq
and similarly for other large- and small-scale fields and their correlations.
We also assume that fluctuating fields are statistically homogeneous, and the forcing functions ${\bm f}^\text{K}$ and ${\bm f}^\text{M}$ are small-scale fields.
We will consider the case with a vanishing background flow $\bar\bmU=\bm 0$ in the present work; including a differentially rotating flow will not introduce significant changes to our method or closure, and will be included in the future work.
Thus our calculations applies to $\bm\alpha^2$ dynamos where anisotropy is due to large-scale magnetic fields or anisotropic forcing, possibly in the presence of a weak differential rotation.

For notational simplicity in the rest of this paper, we will subsequently drop the overlines on mean-field variables (capitalized) but keep them over correlation functions of small-scale fields.
Averaging equation (\ref{eqn:induction0}), we obtain the mean-field induction equation
\beq
\partial_t\bmB=\del\times\left(\bar{\bmu\times\bmb}\right)+\nuM\nabla^2\bmB.
\eeq
The turbulent EMF is defined as $\bm\emf=\bar{\bmu\times\bmb}$.
Due to the statistical homogeneity of small-scale correlations, we have
\beq
\emf_i=\int \frac{\text{d}^3k}{(2\pi)^2}\ 
\epsilon_{ijk}\bar{\tilde u_j^*(\bmk)\tilde b_k(\bmk)}
\label{eqn:emf1}
\eeq
where a tilde indicates a Fourier transform, and the asterisk means complex conjugate.

The turbulent EMF is conventionally expanded in orders of spatial gradients of $\bmB$, i.e.,
\beq
\emf_i=\alpha_{ij}B_j+\beta_{ijk}\partial_j B_k +\mathcal{O}(\kappa^2)
\label{eqn:emf}
\eeq
where $\bm\alpha$ and $\bm\beta$ are tensorial turbulent transport coefficients, and $\kappa$ is the ratio of the turbulent scale to the mean-field scale and keeps track of the order of derivatives of $\bmB$.
Note that for non-vanishing $\bm\emf$, the flow has to be at least weakly isotropic in the sense that $\bar{\bmu\times\bmb}\sim\mathcal{O}(\kappa^0)\neq\bm0$.
Thus our use of ``isotropic transport coefficients'' means that $\bm\alpha$ is isotropic to the $\mathcal{O}(\kappa^0)$ order, $\bm\beta$ to the $\mathcal{O}(\kappa)$ order, etc.

Assuming homogeneity and focusing on the $\bm\alpha$ effect, we may also write
\beq
\emf_i=B_j\int\frac{\text{d}^3k}{(2\pi)^3}\ \tilde\alpha_{ij}(\bmk)
+\mathcal{O}(\kappa),
\label{eqn:emf2}
\eeq
where $\tilde{\bm\alpha}$ is the spectrum of $\bm\alpha$.
To the $\mathcal{O}(\kappa^0)$ order, we can drop gradients of $\bmB$ in expressions for $\bm\alpha$ and $\partial_t\bm\alpha$. From equations (\ref{eqn:NS}) and (\ref{eqn:induction0}) we then obtain the equations for the fluctuating fields to the $\mathcal{O}(\kappa^0)$ order as
\begin{align}
\partial_t\bmu=&\bmB\cdot\del\bmb+\bm Q^\text{K}+\K{\bm f},
\label{eqn:dudt}\\
\partial_t\bmb=&\bmB\cdot\del\bmu+\bm Q^\text{M}+\M{\bm f},
\label{eqn:dbdt}
\end{align}
where we have defined
\begin{align}
\bm Q^\text{K}=&\hat{\bm P}(\bmb\cdot\del\bmb-\bar{\bmb\cdot\del\bmb}
-\bmu\cdot\del\bmu+\bar{\bmu\cdot\del\bmu})+\nu\nabla^2\bmu,\\
\bm Q^\text{M}=&
\del\times(\bmu\times\bmb-\bar{\bmu\times\bmb})+\nuM\nabla^2\bmb
\end{align}
to combine quadratic and diffusive terms.

We next compute the time evolution of the following correlation functions (note that we do not show the explicit time dependence for notational efficiency)
\begin{align}
\K C_{ij}(\bmk)=C_{ji}^{\text{K}*}(\bmk)=&\bar{\tilde u_i^*(\bmk)\tilde u_j(\bmk)},\\
\M C_{ij}(\bmk)=C_{ji}^{\text{M}*}(\bmk)=&\bar{\tilde b_i^*(\bmk)\tilde b_j(\bmk)},\\
\C C_{ij}(\bmk)=&\bar{\tilde u_i^*(\bmk)\tilde b_j(\bmk)}.
\end{align}
To this end, we use the Fourier transforms of equations (\ref{eqn:dudt}) and (\ref{eqn:dbdt}), and obtain
\begin{align}
\partial_t&\K C_{ij}(\bmk)
=i\bmk\cdot\tilde\bmB\left[\C C_{ij}(\bmk)-C_{ji}^{\text{C}*}(\bmk)\right]
+\tilde T_{ij}^\text{K}+\K{\tilde f_{ij}},
\label{eqn:uu}\\
\partial_t&\M C_{ij}(\bmk)
=i\bmk\cdot\tilde\bmB\left[C_{ji}^{\text{C}*}(\bmk)-\C C_{ij}(\bmk)\right]
+\tilde T_{ij}^\text{M}+\M{\tilde f_{ij}},
\label{eqn:bb}\\
\partial_t&\C C_{ij}(\bmk)
=i\bmk\cdot\tilde\bmB\left[\K C_{ij}(\bmk)-\M C_{ij}(\bmk)\right]
+\tilde T_{ij}^\text{C}+\C{\tilde f_{ij}}
\label{eqn:ub},
\end{align}
where
\begin{align}
\tilde T_{ij}^\text{K}=&\bar{\tilde Q_i^\text{K*}(\bmk)\tilde u_j(\bmk)+\tilde u_i^*(\bmk)\tilde Q_j^\text{K}(\bmk)},\\
\tilde T_{ij}^\text{M}=&\bar{\tilde Q_i^\text{M*}(\bmk)\tilde b_j(\bmk)+\tilde b_i^*(\bmk)\tilde Q_j^\text{M}(\bmk)},\\
\tilde T_{ij}^\text{C}=&\bar{\tilde Q_i^\text{K*}(\bmk)\tilde b_j(\bmk)+\tilde u_i^*(\bmk)\tilde Q_j^\text{M}(\bmk)}
\end{align}
are diffusive and triple correlation terms and
\begin{align}
\K{\tilde f_{ij}}=&\bar{\tilde f_i^{\text{K}*}(\bmk)\tilde u_j(\bmk)+\tilde u_i^*(\bmk)\K{\tilde f_j}(\bmk)},\\
\M{\tilde f_{ij}}=&\bar{\tilde f_i^{\text{M}*}(\bmk)\tilde b_j(\bmk)+\tilde b_i^*(\bmk)\M{\tilde f_j}(\bmk)},\\
\C{\tilde f_{ij}}=&\bar{\tilde f_i^{\text{K}*}(\bmk)\tilde b_j(\bmk)+\tilde u_i^*(\bmk)\M{\tilde f_j}(\bmk)}
\end{align}
are forcing correlation terms.
The anisotropy of the correlation functions $C_{ij}^{\text{K,M,C}}$ can either be a result of the factor $\bmk\cdot\bmB$ or come from anisotropic forcing terms $\tilde f_{ij}^{\text{K,M,C}}$.
For future convenience we also define
\beq
\tilde D_{ij}^{\mathcal R}=\tilde T_{ij}^{\mathcal R}
+\tilde f_{ij}^{\mathcal R}
\eeq
for ${\mathcal R}=\text{K, M, C}$. 

The expression for $\bm\alpha$ can now be readily found.
Since $\emf_i=\int\text{d}^3k\ (2\pi)^{-3}\epsilon_{ijk}\C C_{jk}(\bmk)$, in equation (\ref{eqn:ub}) we can apply the minimal-$\tau$ closure to model the sum of the triple correlation term and the forcing term as $-\C C_{ij}(\bmk)/\tau$, where $\tau$ is the $\bmk$-dependent damping time for the cross correlation assumed to be the same for all pairs of $(i,j)$.
For all $\bmk$, $\tau$ is much smaller than the evolution time for mean quantities, including correlation functions, and thus we can drop the time derivative on the left-hand side of equation (\ref{eqn:ub}) to obtain 
\beq
\C C_{ij}(\bmk)
\simeq
(i\bmk\cdot\bmB)
\tau
\left[
\K C_{ij}(\bmk)
-\M C_{ij}(\bmk)
\right].
\label{eqn:ubMTA}
\eeq
Combining equations (\ref{eqn:emf1}), (\ref{eqn:emf2}) and (\ref{eqn:ubMTA}) we see that
\beq
\tilde\alpha_{ij}(\bmk)=i\tau\epsilon_{imn}k_j\left[
\K C_{mn}(\bmk)-\M C_{mn}(\bmk)
\right]=\K{\tilde\alpha}_{ij}
+\M{\tilde\alpha}_{ij},
\eeq
where we have defined
\beq
\tilde{\alpha}^\text{K,M}_{ij}(\bmk)
=\pm i\tau\epsilon_{imn}k_j C^\text{K,M}_{mn}(\bmk).
\label{eqn:tilde_alpha_KM}
\eeq
Correspondingly,
\beq
\emf_i=B_j\int \frac{\text{d}^3k}{(2\pi)^3}\ 
\left[
\K{\tilde\alpha}_{ij}(\bmk)
+\M{\tilde\alpha}_{ij}(\bmk)
\right]
+\cdots.
\label{eqn:emf3}
\eeq

{Note that in equation (\ref{eqn:tilde_alpha_KM}), the $\M C_{mn}$ term comes from the cross product of $\bmb$ and the Lorentz force \ in the Navier-Stokes equation.
For the isotropic case, 
\beq
k_j C^\text{K,M}_{mn}(\bmk)
=\frac{1}{6}\epsilon_{jmn}
\epsilon_{ikl}
k_i C^\text{K,M}_{kl}(k),
\label{eqn:isotropization}
\eeq
so that 
 $k_j C^\text{K,M}_{mn}$ is completely
determined by the helicity terms $\epsilon_{ikl}k_i C^\text{K,M}_{kl}$, and thus using helicity conservation is equivalent to including the full Lorentz back-reaction.
But for general anisotropic cases, equation (\ref{eqn:isotropization}) no longer holds, and using helicity conservation alone excludes the anisotropic part of the $k_j C^\text{K,M}_{mn}$ tensor, thereby failing to account for the full Lorentz back-reaction }

In either isotropic or anisotropic cases, the elements of $\tilde{\bm\alpha}^\text{K,M}$ are all proportional to its trace.
To see this, note that $\epsilon_{ijk}\tilde{\alpha}^\text{K,M}_{jk}=0$ for incompressible flows, and therefore
\beq
0=\epsilon_{iab}\epsilon_{ijk}\tilde{\alpha}^\text{K,M}_{jk}
=\tilde{\alpha}^\text{K,M}_{ab}-\tilde{\alpha}^\text{K,M}_{ba}.
\eeq
This implies that $\tilde{\bm\alpha}^\text{K,M}$ are symmetric tensors, or
\beq
i\tau\epsilon_{imn}k_j C^\text{K,M}_{mn}
=i\tau\epsilon_{jmn}k_i C^\text{K,M}_{mn}.
\eeq
Multiplying both sides by $k_j$ and summing over $j$ we obtain
\beq
\epsilon_{imn} k^2 C^\text{K,M}_{mn}
=i\tau \epsilon_{jmn} k_i k_j C^\text{K,M}_{mn}=\pm k_i\tr\tilde{\bm\alpha}^\text{K,M},
\eeq
or $\epsilon_{imn}C^\text{K,M}_{mn}=\pm k_i\tr\tilde{\bm\alpha}^\text{K,M}/k^2$, and henceforth
\beq
\tilde{\bm\alpha}^\text{K,M}=\hat k_{ij}\tr\tilde{\bm\alpha}^\text{K,M}
\equiv \hat k_{ij} \tilde\alpha^\text{K,M}
\label{eqn:trace_form}
\eeq
where $\hat k_{ij}=k_ik_j/k^2$, $k=|\bmk|$, and we have used unbold variables to denote their traces.
Equation (\ref{eqn:trace_form}) implies that the time evolution of the spectra of the kinetic and magnetic $\bm\alpha$ tensors are solely determined by their traces, whereby
\newcommand{\KM}[1]{{#1}^\text{K,M}}
\beq
\KM\alpha_{ij}(t)
=\int\frac{\text{d}^3k}{(2\pi)^3}\ 
\KM{\tilde\alpha}_{ij}(t,\bmk)
=\int\frac{\text{d}^3k}{(2\pi)^3}\ 
\hat k_{ij} \KM{\tilde\alpha}(t,\bmk).
\label{eqn:trace_form2}
\eeq

\subsection{New equations for time evolution of $\bm\alpha$}
Substituting the expression for $\C C_{ij}$ [equation (\ref{eqn:ubMTA})] into equations (\ref{eqn:uu}) and (\ref{eqn:bb}) we have
\beq
\partial_t \K C_{ij}(\bmk)
=-2\tau(\bmk\cdot\bmB)^2\left[\K C_{ij}(\bmk)-\M C_{ij}(\bmk)\right]
+\tilde D_{ij}^\text{K},
\eeq
and
\beq
\partial_t\M C_{ij}(\bmk)=2\tau
(\bmk\cdot\bmB)^2\left[\K C_{ij}(\bmk)-\M C_{ij}(\bmk)\right]
+\tilde D_{ij}^\text{M}.
\eeq
Then using equation (\ref{eqn:tilde_alpha_KM}) we have
\beq
\partial_t\KM{\tilde\alpha}_{ij}
=-2\tau(\bmk\cdot\bmB)^2\left(
\K{\tilde\alpha}_{ij}+\M{\tilde\alpha}_{ij}
\right)+\KM{\tilde\xi}_{ij}.
\label{eqn:ddt_full}
\eeq
where
\beq
\KM{\tilde\xi}_{ij}(\bmk)
=\pm i\tau\epsilon_{imn}k_j
\KM{\tilde D}_{mn}.
\eeq

Using equation (\ref{eqn:trace_form}), equation (\ref{eqn:ddt_full}) can be further reduced to
\beq
\partial_t\KM{\tilde\alpha}
=-2\tau(\bmk\cdot\bmB)^2\left(
\K{\tilde\alpha}+\M{\tilde\alpha}\right)
+\KM{\tilde\xi}
\label{eqn:ddt_alpha_km}
\eeq
where $\KM{\tilde\xi}=\tr\KM{\tilde{\bm\xi}}$ includes the combined effects of forcing, turbulent cascade, and microscopic diffusivity and resistivity. Here we adopt a ``selective-damping-$\tau$'' (SDT) closure which models $\KM{\tilde\xi}$ using physical arguments by considering their evolution in the absence of the mean field $\bmB$.
This SDT closure is a new variation of the 
$\tau$ approximation which, as in previous related closures \citep{Pouquet1976,BlackmanField2002prl,Rogachevskii2003}, 
approximates third-order moments by a restoring term in the equations of second-order moments.
But here we further impose on the restoring terms constraints from conservation laws of the system. This influences
the relative damping strength of the closure for different second-order moments, and thus we refer to this closure as one that produces ``selective'' damping.

More explicitly, we will consider a kinetically forced flow and we first consider $\K{\tilde\xi}$.
As the turbulence is kinetically forced, it is natural to expect that $\K{\tilde{\xi}}$ drives $\K{\tilde{\alpha}}$ back to its constant background forced value $\tilde{\alpha}^0$ on a forcing time, which equals the eddy time scale $\tau$.
 We thus adopt
\newcommand{\kL}{{k_\text{L}}}
\beq
\K{\tilde{\xi}}\simeq
-\frac{\K{\tilde{\alpha}}-\tilde{\alpha}^0}{\tau}.
\label{eqn:xik}
\eeq
We also assume that the forcing is stronger than the small-scale magnetic back-reaction at all times and thus the latter contributes negligibly to $\K{\tilde{\xi}}$.

For $\M{\tilde{\xi}}$, we expect a somewhat different form from equation (\ref{eqn:xik}) as there is no external magnetic forcing and $\bmb$ is driven by a small-scale dynamo. Our form of $\M{\tilde\xi}$ will be constructed using the following physical arguments:

(i) Note that $\M{\tilde\alpha}/(\tau k^2)$ is the small-scale magnetic helicity spectrum and exhibits an inverse cascade.
An inverse transfer from a turbulent scale towards a mean-field scale would occur on a time scale $\kL/b\gtrsim\kL/u$, much longer than $(\tau \ks^2 B^2)^{-1}\sim \ks/u$ where $\kL$ and $\ks$ are respectively the scales of mean fields and turbulent fields.
Any inverse transfer within the turbulent inertial range, e.g., from a smaller scale towards the outer or forcing scale
 vanishes after integration over $\bmk$.
We thus require that when $\bmB=\bm 0$, the small-scale magnetic helicity is conserved.
In terms of the evolution of $\M{\tilde\alpha}$,
\beq
\partial_t\bar{\bm a\cdot\bmb}
=\partial_t\int\frac{\text{d}^3 k}{(2\pi)^3}\ 
\frac{\M{\tilde\alpha}}{\tau k^2}
=\int\frac{\text{d}^3 k}{(2\pi)^3}\ 
\frac{\M{\tilde\xi}}{\tau^2 k^2}
=0.
\label{eqn:xi_constraint1}
\eeq

(ii) To $\mathcal{O}(\kappa^0)$, the dynamo would saturate
when $\K{\tilde\alpha}+\M{\tilde\alpha}= 0$, 
and we thus require that
\beq
\M{\tilde\xi}|_{\K{\tilde\alpha}=-\M{\tilde\alpha}}
=0.
\label{eqn:xi_constraint2}
\eeq

(iii) In a saturated small-scale dynamo, $\bmb$ is largely parallel or anti-parallel to the vorticity field $\del\times\bmu$ due to the formal similarity of their equations of motion \cite{Brandenburg1996, BrandenburgSubramanian2005}.
Note that if $\bmb{|_\text{sat}}\propto\del\times\bmu{|_\text{sat}}$, we would have $\M{\tilde\alpha}{|_\text{sat}}\propto k^2\K{\tilde\alpha}{|_\text{sat}}$.
Additionally motivated by equation (\ref{eqn:xi_constraint2}), we assume the more general relation $\M{\tilde\alpha}{|_\text{sat}}=\zeta \K{\tilde\alpha}{|_\text{sat}}$ where $\zeta$ is some function of $\bmk$.
For single-scale turbulence, $\M{\tilde\alpha}=\zeta \K{\tilde\alpha}$ is reduced to $\bmb\propto\del\times\bm u$, while for broad-spectrum turbulence the latter is approximately true when the energy has a dominant scale.
The $\M{\tilde\xi}$ term is then assumed to drive the helical part of $\bmb$ to the saturated state in the absence of $\bmB$ within some time scale $\tau_\text{ssd}$, i.e.,
\beq
\M{\tilde\xi}
=-\frac{\M{\tilde\alpha}-\zeta(\bmk)\K{\tilde\alpha}}{\tau_\text{ssd}}.
\label{eqn:xi_constraint3}
\eeq

(iv) 
Physically, while $\M{\tilde\xi}$ conserves the small-scale magnetic helicity in the absence of $\bmB$, it modifies the geometry of the field lines by adjusting the distribution of magnetic helicity at different directions and scales.
For stronger helical magnetic fields, changing its helicity spectrum presumably requires more energy, and thus for a given forcing, it takes a longer time to do so.
We thus assume that the driving time scale $\tau_\text{ssd}$ is larger when a strongly helical field $\bmb$ is driven by a weakly helical $\bmu$.

\newcommand{\kf}{{k_\text{f}}}
Taking all four above constraints into consideration, we propose the following SDT closure for the nonlinear terms in $\partial_t\M{\tilde\alpha}$:
\beq
\M{\tilde\xi}
=-\frac
{\M{\tilde\alpha}-\zeta\K{\tilde\alpha}}
{e^{\zeta^2}\tau}
\label{eqn:xiM}
\eeq
where
\beq
\zeta=\frac
{\int\text{d}^3k\ 
\M{\tilde\alpha}/(\tau^2 k^2)}
{\int\text{d}^3k\ \K{\tilde\alpha}/(\tau^2 k^2)},
\label{eqn:zeta}
\eeq
and the driving time scale $\tau_\text{ssd}=e^{\zeta^2}\tau$ becomes large when $\bmb$ is more helical than $\bmu$.
The detailed shape factor ($=\tau_\text{ssd}/\tau$) is presumably not important as long as $\tau_\text{ssd}\to\tau$ as $\zeta\to0$ and $\tau_\text{ssd}\to\infty$ as $\zeta\to\infty$.
Also note that $\M{\tilde\xi}\to0$ does not imply that $\bmb$ is not amplified since $\M{\tilde\alpha}$ only involves the anti-symmetric, non-energy-carrying part of $\bar{b_i b_j}$.


Applying the SDT closure to equation (\ref{eqn:ddt_alpha_km}) we arrive at
\begin{align}
\partial_t\K{\tilde{\alpha}}
=&-2\tau(\bmk\cdot\bmB)^2\tilde{\alpha}
-\frac{\K{\tilde{\alpha}}-\tilde{\alpha}^0}{\tau},
\label{eqn:ddt_alpha_k}\\
\partial_t\M{\tilde{\alpha}}
=&-2\tau(\bmk\cdot\bmB)^2\tilde{\alpha}
-\frac
{\M{\tilde\alpha}-\zeta\K{\tilde\alpha}}
{e^{\zeta^2}\tau},
\label{eqn:ddt_alpha_m}
\end{align}
where $\tilde\alpha=\K{\tilde\alpha}+\M{\tilde\alpha}$.
We note that these equations conserve the total magnetic helicity to the $\mathcal{O}(\kappa^0)$ order:
\beq
\partial_t\abra{\int \frac{\text{d}^3k}{(2\pi)^3}
\frac{\M{\tilde\alpha}}{\tau k^2}}
=-2\abra{\alpha_{ij}B_i B_j}
=-\partial_t\abra{\bm A\cdot\bmB}
\eeq
where angle brackets indicate a volume average.

Importantly, the SDT quenching includes the complete Lorentz back-reaction and thus can not only fully offset the $\bm\alpha$ tensor, but also captures the influence of the small-scale dynamo. This generally means faster-than-conventional quenching because $\bmb$ tends to align with $\bmu$ on the $\tau_\text{ssd}$ time scale when the full Lorentz force is included.

Equations (\ref{eqn:ddt_alpha_k}) and (\ref{eqn:ddt_alpha_m}) are the most fundamental theoretical result of this paper, 
with subsequent sections exploring their limits, or solving them with examples.

\subsection{Comparison with conventional quenching for isotropic, single-scale turbulence}
For conventional isotropic turbulence, spectra of fluctuating fields are assumed to strongly peak at the energy-dominant scale $\ks$.
In addition, we use the isotropic condition (\ref{eqn:isotropization}), 
which yields the turbulent EMF
\newcommand{\taus}{\tau_\text{S}}
\beq
\emf_i=\frac{\taus}{3}\left(
-\bar{\bmu\cdot\del\times\bmu}
+\bar{\bmb\cdot\del\times\bmb}
\right)B_i
=\frac{1}{3}\alpha B_i,
\eeq
where $\taus=\tau(\ks)$.

In conventional minimalist approaches, the kinetic helicity, $\bar{\bmu\cdot\del\times\bmu}$, is commonly assumed to be a constant, steadily supplied by external forcing (e.g. convection in stars, or supernova explosions in galaxies).
The evolution of $\alpha$ is then solely determined by the evolution of current helicity, $\bar{\bmb\cdot\del\times\bmb}$, which is proportional to the large-scale magnetic helicity through helicity conservation, i.e., using $\bar{\bmb\cdot\del\times\bmb}=\ks^2\bar{\bm a\cdot\bmb}$ in a single-scale turbulence and
\beq
\partial_t\left(\abra{\bm A\cdot\bmB}+\bar{\bm a \cdot\bmb}\right)=0.
\label{eqn:helicity_conservation}
\eeq
The quenching of $\alpha$ is then determined by 
\beq
\partial_t\alpha
=\taus \partial_t\left(\bar{\bmb\cdot\del\times\bmb}\right)
\simeq -\taus k_\text{S}^2\partial_t\abra{\bm A\cdot\bmB}
\simeq -\frac{2}{3}\taus k_\text{S}^2\abra{B^2}\alpha,
\label{eqn:quenchingIso}
\eeq
where in the last equality we have used the equation of $\partial_t(\bm A\cdot\bmB)$ for an $\alpha^2$ dynamo.

The key ingredients to derive the quenching formula (\ref{eqn:quenchingIso}) are (i) a single-scale turbulence (i.e., a two-scale approach) which makes $\bar{\bmb\cdot\del\times\bmb}$ proportional to $\bar{\bm a\cdot\bmb}$, and (ii) magnetic helicity conservation (\ref{eqn:helicity_conservation}) which connects the time evolution of small-scale fields to mean fields.
For a general anisotropic flow, neither of these assumptions hold:
(i) The turbulence is a multi-scale phenomenon 
and
(ii) in the general case (\ref{eqn:emf3}) there is no ``shortcut'' like using helicity conservation to link $\M{\bm \alpha}$ to mean fields.
One must then use the full evolution equations (\ref{eqn:uu}) to (\ref{eqn:ub}) to derive a general quenching formula, i.e., equations (\ref{eqn:ddt_alpha_k}) and (\ref{eqn:ddt_alpha_m}), which are valid for arbitarily anisotropic flows.

To derive the two-scale isotropic limit of SDT quenching, we use $\KM{\tilde{\alpha}}_{ij}(\bmk)=\KM{\tilde{\alpha}}_{ij}(k)\propto\delta_{ij}\delta(k-\ks)$.
Integrating equations (\ref{eqn:ddt_alpha_k}) and (\ref{eqn:ddt_alpha_m}) over $\bmk$ then gives
\begin{align}
\partial_t\K{\alpha}
=&-\frac{2\taus\ks^2 B^2}{3}\alpha
-\frac{\K{\alpha}-\alpha^0}{\taus},
\label{eqn:ddt_tr_k}\\
\partial_t\M{\alpha}
=&-\frac{2\taus\ks^2 B^2}{3}\alpha,
\label{eqn:ddt_tr_m}
\end{align}
where unbold variables indicate their traces.
Also note that the second term on the right side of equation (\ref{eqn:ddt_alpha_m}) correctly vanishes in this single-scale formalism because $\M\alpha$ is now proportional to the small-scale magnetic helicity and hence has to be conserved when $\bmB= 0$.
For the sake of comparison, we write the conventional quenching formula 
\citep{BlackmanField2002prl, BrandenburgSubramanian2005} as
\begin{align}
\partial_t\K{\alpha}=&0,
\label{eqn:ddt_tr_ak_conv}\\
\partial_t\M{\alpha}=&
-\frac{2\taus\ks^2\abra{B^2}}{3}\alpha.
\label{eqn:ddt_tr_am_conv}
\end{align}

Equations (\ref{eqn:ddt_tr_k}) and (\ref{eqn:ddt_tr_m}) are nearly these conventional quenching formula except that:

(i) The SDT quenching depends on the local values of $\bmB$ whereas in the conventional quenching depends on the volume-averaged value $\abra{B^2}$ because magnetic helicity is only conserved globally.
Our new approach is thus physically more sensible in systems whose Alfv\'en crossing time is larger than the turbulent eddy turnover time, which is the case for most astrophysical systems like galaxies and stars.

(ii) The conventional quenching assumes a constant kinetic helicity determined by $\alpha^0$.
In our new formalism this is established by a restoring term $-(\K{\alpha}-\alpha^0)/\tau$ which is capable of capturing its dynamics.

(iii) The quenching of $\K{\alpha}$ is also novel in our new approach.
A time-dependent $\bm\alphaK$ is consistent with numerical results (e.g., Ref. \cite{Brandenburg2001}).
We leave further implications of $\K{\bm\alpha}$ saturation (e.g. its dependence on magnetic Prandtl number) for future work.

\section{Examples of $\bm\alpha^2$ dynamos}
\label{sec:3}
In this section we first perform a modal analysis and show that a mean-field $\bm\alpha^2$ dynamo does exist in an anisotropic and even non-helical flow (i.e., with $\tr\bm\alpha=0$).
We then compare numerical results of solving three different mean-field $\bm\alpha^2$ dynamos with different choices of $\bm\alpha^0$ (isotropic, traceful anisotropic, or traceless anisotropic) and compare different quenching prescriptions for these choices (conventional vs. SDT). 

\subsection{Modal analysis}
We have shown that for incompressible flows $\bm\alpha$ is symmetric according to equation (\ref{eqn:trace_form2}).
We then move to a frame where the local mean flow $\bmU$ vanishes.
To find necessary conditions for growthful large-scale magnetic energy, it is sufficient to first consider only kinetic $\bm\alpha$ in the absence of diffusion, i.e., we will consider an imposed $\bm\alpha$ supplied by flow helicity.
In this case,
\beq
\partial_t B_i=\epsilon_{ijk}\partial_j\alpha_{kl} B_l,
\label{eqn:induction}
\eeq
or, in Fourier space,
\beq
\partial_t\tilde B_i(t,\bmk)
=i\epsilon_{imn}k_m\alpha_{nj}\tilde B_j
\equiv M_{ij}(\bmk)\tilde B_j.
\label{eqn:induction2}
\eeq
The formal solution of equation (\ref{eqn:induction2}) can be written as
\beq
\tilde\bmB(t,\bmk)
=e^{\bm M(\bmk)t}\tilde\bmB(0,\bmk)
\equiv\bmcalU(t,\bmk)\tilde\bmB(0,\bmk).
\label{eqn:formalsolution}
\eeq

The matrix coefficient $M_{ij}$ in equation (\ref{eqn:induction2}) is in general non-normal, i.e., $\bm M \bm M^\dagger\neq\bm M^\dagger \bm M$, where $\bm M^\dagger$ is the Hermitian transpose of $\bm M$, $\left(\bm M^\dagger\right)_{ij}=M_{ji}^*$.
Since the eigenvectors of a non-normal matrix are not orthogonal (see, e.g., Ref. \cite{Trefethen1993}), their
individual growth or decay at early times may not be representantive of the total energy growth at late times. 
Nevertheless, at late times [$t\gtrsim (\alpha_\text{max} k_\text{L})^{-1}$ where $k_\text{L}$ the the scale of the mean fields, and $\alpha_\text{max}=\text{max}\left\{|\lambda_1|,|\lambda_2|,|\lambda_3|\right\}$ with $\lambda_{1,2,3}$ being the eigenvalues of $\bm\alpha$], $\tilde\bmB$ is still dominated by the mode 
with the largest growth rate. 

We now analyze the eigenvalues of the evolution operator $M_{ij}=i\epsilon_{imn}k_m\alpha_{nj}$.
In the frame whose basis vectors are parallel to the principal axes of $\bm\alpha$ (and thus $\bm\alpha=\text{diag}\{\lambda_1,\lambda_2,\lambda_3\}$), one of the eigenvalues of $M_{ij}$
is zero,
and the other two are \citep{Moffatt1978, Rasskazov2018pre}
\beq
\gamma_\pm=\pm\sqrt{
	k_1^2\lambda_{2}\lambda_{3}+k_2^2\lambda_{3}\lambda_{1}+k_3^2\lambda_{1}\lambda_{2}
}.
\label{eqn:lambdaPM2}
\eeq

The existence of growing modes of (\ref{eqn:lambdaPM2}) was shown in Sec 9.3 of Ref. \cite{Moffatt1978}, but there the subsequent discussion focused on cases of helical flows.
Below we explain how non-helical (traceless $\bm\alpha$) flows admit growing modes.
Such traceless $\bm\alpha$ modes can have at most one vanishing eigenvalue.
When $\bm\alpha$ has $0$ as an eigenvalue, let $\bm\alpha=\text{diag}\{\lambda,-\lambda,0\}$ with $\lambda>0$.
The corresponding eigenvalues of $M_{ij}$ 
are $\gamma_\pm=\pm i \lambda k_3$, and the sum of such temporally periodic Fourier modes need not necessarily yield a temporally periodic function \footnote{For example, consider a function $f(t,x)$ whose Fourier transform with respect to $x$ is $e^{i\nu k t}f_1(k)$ where $\nu$ is a constant.
Then the Fourier transform of $f$ with respect to $t$ is $f_2(\omega,x)=\int\text{d}t\ e^{-i\omega t}\int\text{d}k\ e^{ikx} e^{i\nu k t}f_1(k)=e^{i\omega x/\nu}f_1(\omega/\nu)$ which can have a continuous spectrum, implying that $f(t,x)$ can be non-periodic in $t$.}.
In fact an analytical solution of equation (\ref{eqn:induction2}) is available with such an $\bm\alpha$ and an isotropic initial condition for $\bmB$
that grows large-scale magnetic energy, albeit more slowly than exponential.


If instead $\bm\alpha$ has no vanishing eigenvalue, then without loss of generality let $\bm\alpha=\text{diag}\{\lambda_1,\lambda_2,-\lambda_1-\lambda_2\}$ with $\lambda_1\lambda_2>0$.
Equation (\ref{eqn:lambdaPM2}) then becomes
\beq
\gamma_\pm=
\pm i\sqrt{(\lambda_2k_1)^2+(\lambda_1k_2)^2+\lambda_1\lambda_2(k_1^2+k_2^2-k_3^2)},
\label{eqn:gamma}
\eeq
which has real and positive value for example, when $k_1=k_2=0$ and the magnetic fields are averaged over the $xy-$plane. The growth rate is then $\sqrt{\lambda_1\lambda_2 k_3^2}$.
Generally, the condition for $\gamma_+$ to be positive is
\beq
k_1^2+\chi k_2^2-\frac{\chi}{1+\chi}k_3^2<0,
\label{eqn:wavenumberCondition}
\eeq
with $\chi=\lambda_1/\lambda_2>0$.
Provided that $k_1^2$ and $k_2^2$ are sufficiently small, growing modes always exist.
This shows that a non-helical flow can generate large-scale magnetic fields through the anisotropic $\bm\alpha$ effect.

\subsection{Case of isotropic single-scale $\bm\alpha$}
In this subsection and the two that follow, we demonstrate numerically how an $\bm\alpha^2$ dynamo saturates with the conventional vs. the SDT quenching formulae.
To this end, we solve the evolution of the mean-field vector potential
\beq
\partial_t A_i=\alpha_{ij}\left\{\del\times\bm A\right\}_j
\label{eqn:induction_A}
\eeq
and compute the corresponding magnetic field $\bmB=\del\times\bm A$ and the spatial average $\abra{B^2}$.
The quenching recipies we use are
\beq
\bm\alpha(t)=\bm\alpha(0)-\frac{1}{3}\taus\ks^2\abra{\bm A\cdot\bmB}\bm I_3
\eeq
for the conventional one, and equations (\ref{eqn:ddt_alpha_k}) and (\ref{eqn:ddt_alpha_m}) for SDT quenching.

\newcommand{\bmkL}{{\bmk_\text{L}}}
\newcommand{\kLx}{{k_\text{L1}}}
\newcommand{\kLy}{{k_\text{L2}}}
\newcommand{\kLz}{{k_\text{L3}}}
\newcommand{\urms}{{u_\text{rms}}}
We normalize velocity and magnetic fields by $\urms=\left(\bar{u^2}\right)^{1/2}$, wavenumbers by $\kL$, and time by $(\urms\kL)^{-1}$.
In these units we choose a non-helical initial condition $\bm A(0)=\bm A_0\cos\left(\bmkL\cdot\bm x\right)$ where $\bm A_0=10^{-4}\times\left(4,5,6\right)$ is a constant vector, and $\bmkL=(\kLx,\kLy,\kLz)$ is a constant unit vector chosen such that the dynamo has at least one growing eigenmode according to equation (\ref{eqn:wavenumberCondition}).
The equations are solved in a periodic box $\bm x\in [0,2\pi/\kLx]\times[0,2\pi/\kLy]\times[0,2\pi/\kLz]$.

For the single-scale isotropic case we choose $\bmkL=(1,1,1)/\sqrt{3}$, and adopt a constant time scale $\taus=0.1$ and wave number $\ks=10$ for the turbulent fields.
Also note that in this case, equations (\ref{eqn:ddt_alpha_k}) and (\ref{eqn:ddt_alpha_m}) reduce to equations (\ref{eqn:ddt_tr_k}) and (\ref{eqn:ddt_tr_m}).
We use the initial conditions
\beq
\K{\bm\alpha}(0)=\bm\alpha^0=\bm I_3,\ \M{\bm\alpha}(0)=\bm 0.
\eeq

\begin{figure}
\centering
\includegraphics[width=0.8\columnwidth]{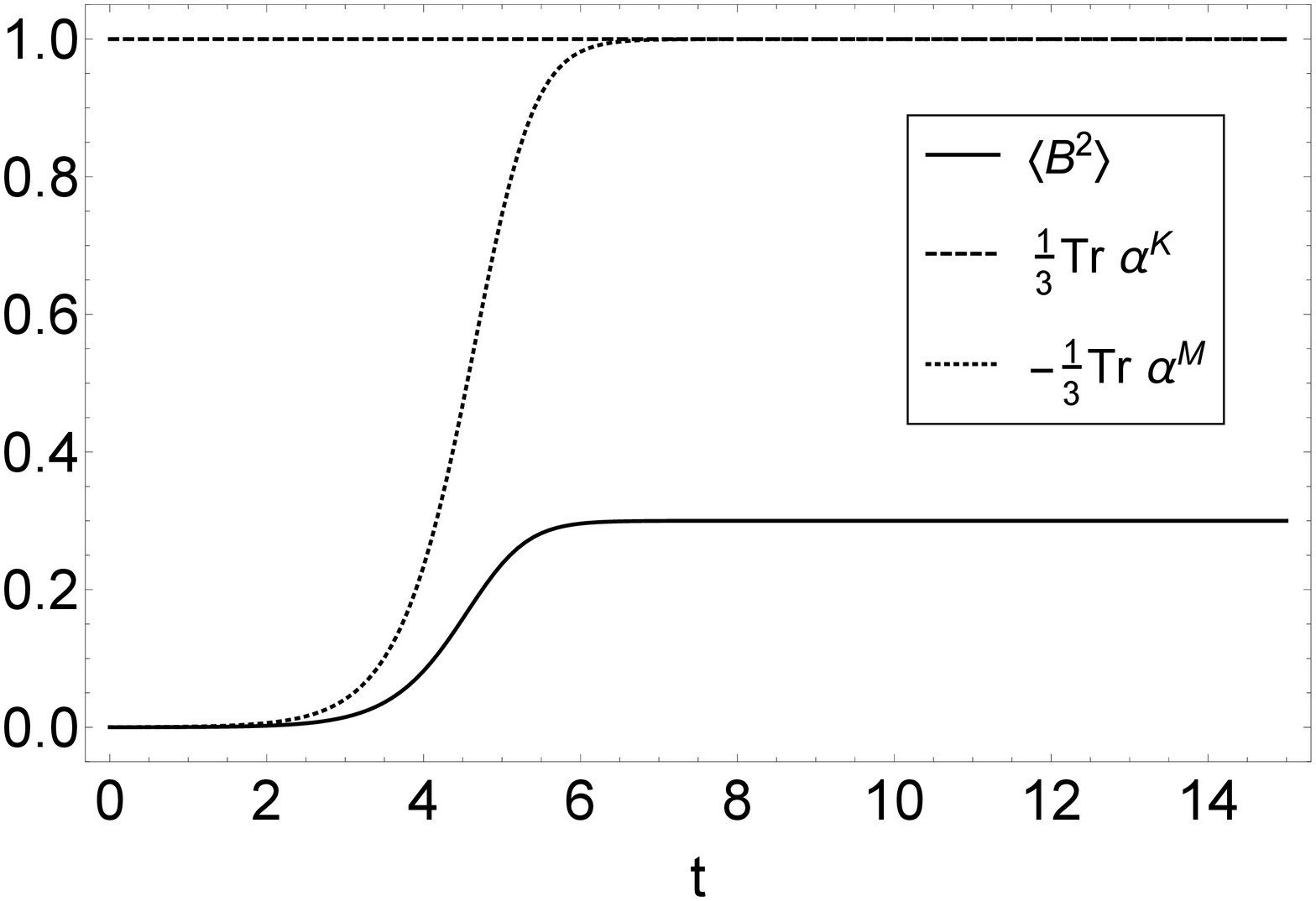}\\
\includegraphics[width=0.8\columnwidth]{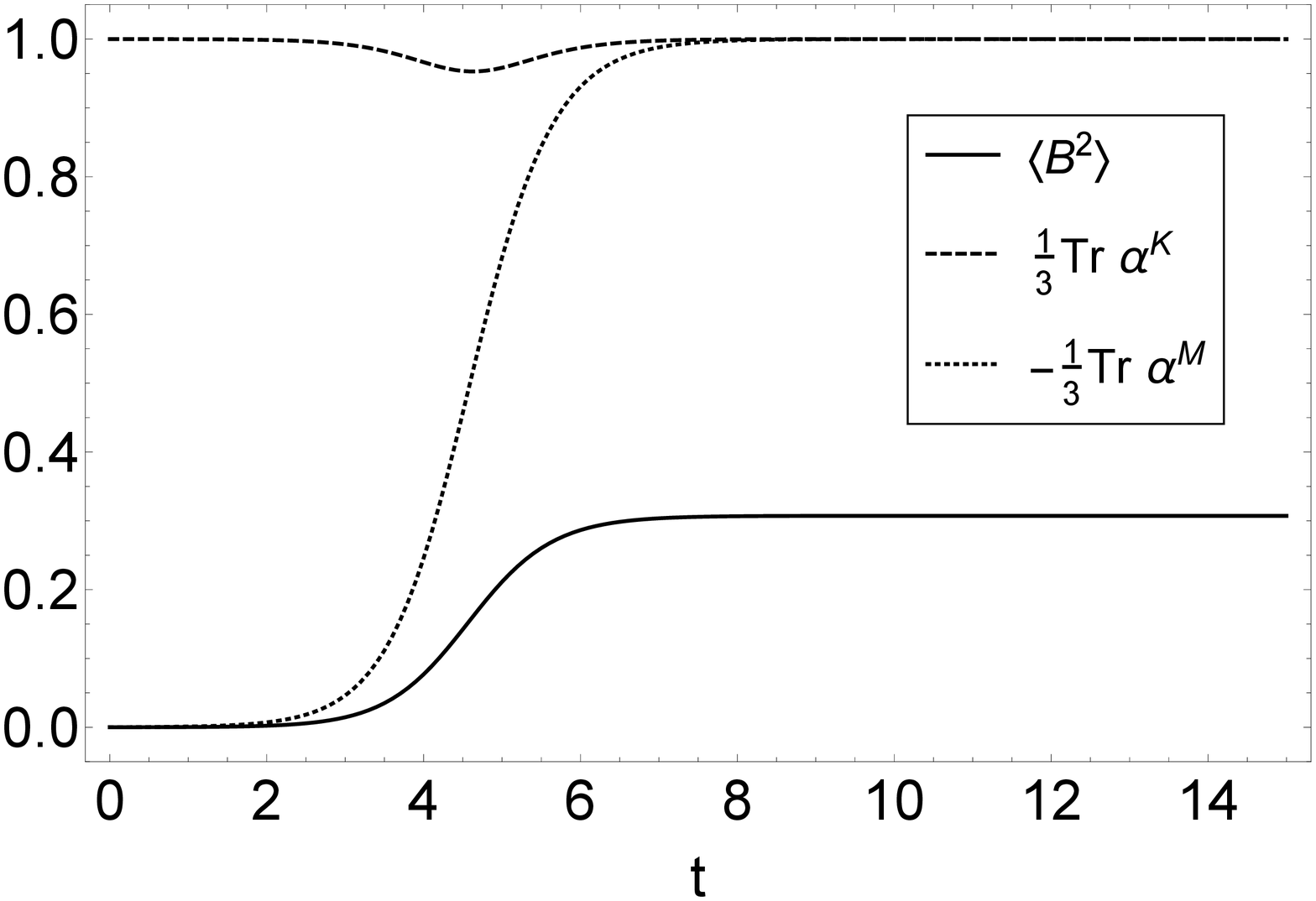}
\caption{Time evolution of $\abra{B^2}$ (solid), $\tr\K{\bm\alpha}/3$ (dashed), and $-\tr\M{\bm\alpha}/3$ (dotted) in the isotropic single-scale turbulence case with the conventional quenching (top) and the SDT quenching (bottom).
The vertical axis is normalized by $\abra{u^2}$, and $t$ is measured in $\sqrt{\abra{u^2}\kL^2}$.
}
\label{fig:iso}
\end{figure}

The results are shown in figure \ref{fig:iso}.
The two quenching formalisms give almost identical results for the evolution of the total magnetic energy ($\propto\abra{B^2}$), small-scale kinetic helicity ($\propto-\tr\K{\bm\alpha}$), and small-scale current helicity ($\propto\tr\M{\bm\alpha}$), except that $\tr\K{\bm\alpha}$ exhibits a small drop and then quickly recovers it background value before the saturation.
Not surprisingly, the saturated value for the mean-field magnetic energy can be estimated by
\begin{align}
&\abra{\bm A\cdot\bmB}=-\bar{\bm a\cdot\bmb}\notag\\
\Rightarrow&\frac{\abra{B^2}}{\kL}\simeq
-\frac{\bar{\bmb\cdot\del\times\bmb}}{\ks^2}
=-\frac{\tr\bm\alpha^0}{\taus\ks^2}\notag\\
\Rightarrow&\abra{B^2}_\text{sat}\simeq
-\frac{\kL\tr\bm\alpha^0}{\taus\ks^2}
\simeq0.3
\end{align}
using magnetic helicity conservation.

\subsection{Case of anisotropic traceful $\bm\alpha^0$}
\newcommand{\knu}{{k_\nu}}
For anisotropic flows, the SDT quenching describes the evolution of the spectra of $\K{\bm\alpha}$ and $\M{\bm\alpha}$.
To illustrate this, we adopt a Kolmogorov spectrum for the turbulence with an inertial range $\kf\leq k\leq \knu$, where $\kf=\ks=10$ and $\knu=100$ are the wave numbers at the forcing scale and the dissipation scale (assume unity magnetic Prandtl number), respectively.
We then have $\tau(k)=\tau_0 (k/\kf)^{-2/3}$ and choose $\tau_0=\taus=0.1$.
For the initial spectrum of the turbulent velocity field, we assign the following helical spectrum of $\K{\bm C}(0)$:
\beq
\K C_{ij}(0)=
P_{ij}(\bmk)F_1(\bmk)
-ik_l\epsilon_{lij}\left(1-\frac{k_3^2}{k^2}\right)
\frac{15\pi^2F_2(k)}{4k^4},
\label{eqn:aniso_spectrum}
\eeq
with $F_2(k)=f_2 (k/\kf)^{-2/3}$, $f_2$ being a constant, and the imposed normalization $\int\text{d}k\ \tau(k) F_2(k)=3\tau_0 f_2\kf[1-(\kf/\knu)^{1/3}]=1$ yielding $f_2\simeq0.62$.
Using equation (\ref{eqn:tilde_alpha_KM}) and integrating $\tilde{\bm\alpha}^0=\K{\tilde{\bm\alpha}}(0)$ over $\bmk$ gives
\beq
\bm\alpha^0=\text{diag}\left\{
1,1,\frac{1}{2}
\right\}.
\label{eqn:initial_alpha_helical}
\eeq
This form of $\bm\alpha$ has the same symmetry as that of the $\bm\alpha$ tensor in a stratified rotating turbulence whose directions of stratification and rotation are parallel, e.g., in galactic dynamos \citep{Raedler2003,Brandenburg2013,ZhouBlackman2019}.
The anisotropic factor $(1-k_3^2/k^2)$ in equation (\ref{eqn:aniso_spectrum}) can be understood as the following.
A radially differentially rotating flow that has no velocity gradient in the $\hat{\bm z}$ direction, should have vanishing spectrum when $\hat{\bm z}\cdot\hat{\bmk}=1$.
In addition, this anisotropic factor should be even under spatial inversion because $F_2\propto \bm\Omega\cdot\hat{\bm z}$ is odd under spatial inversion and as a pseudo-scalar $\tr\bm\alpha$ has to be odd too.
Thus $(1-k_3^2/k^2)$ is a natural choice.

For the SDT quenching we thus have an initial value problem in $\bmk$ space:
Equations (\ref{eqn:ddt_alpha_k}) and (\ref{eqn:ddt_alpha_m}), with the initial condition
\beq
\K{\tilde{\alpha}}(0)=
\tilde\alpha^0=
\left(1-\frac{k_3^2}{k^2}\right)
\frac{15\pi^2 \tau F_2(k)}{2k^2},\ 
\M{\tilde{\alpha}}(0)=0.
\eeq
For approximately homogeneous $\bm\alpha$ with the assumption of large scale separation, we use $(\bmk\cdot\bmB)^2\simeq k_i k_j\abra{B_iB_j}$.
For the initial condition of $\bm A$, we choose $\bmkL=(5,5,1)/\sqrt{51}$.

\begin{figure}
\centering
\includegraphics[width=0.8\columnwidth]{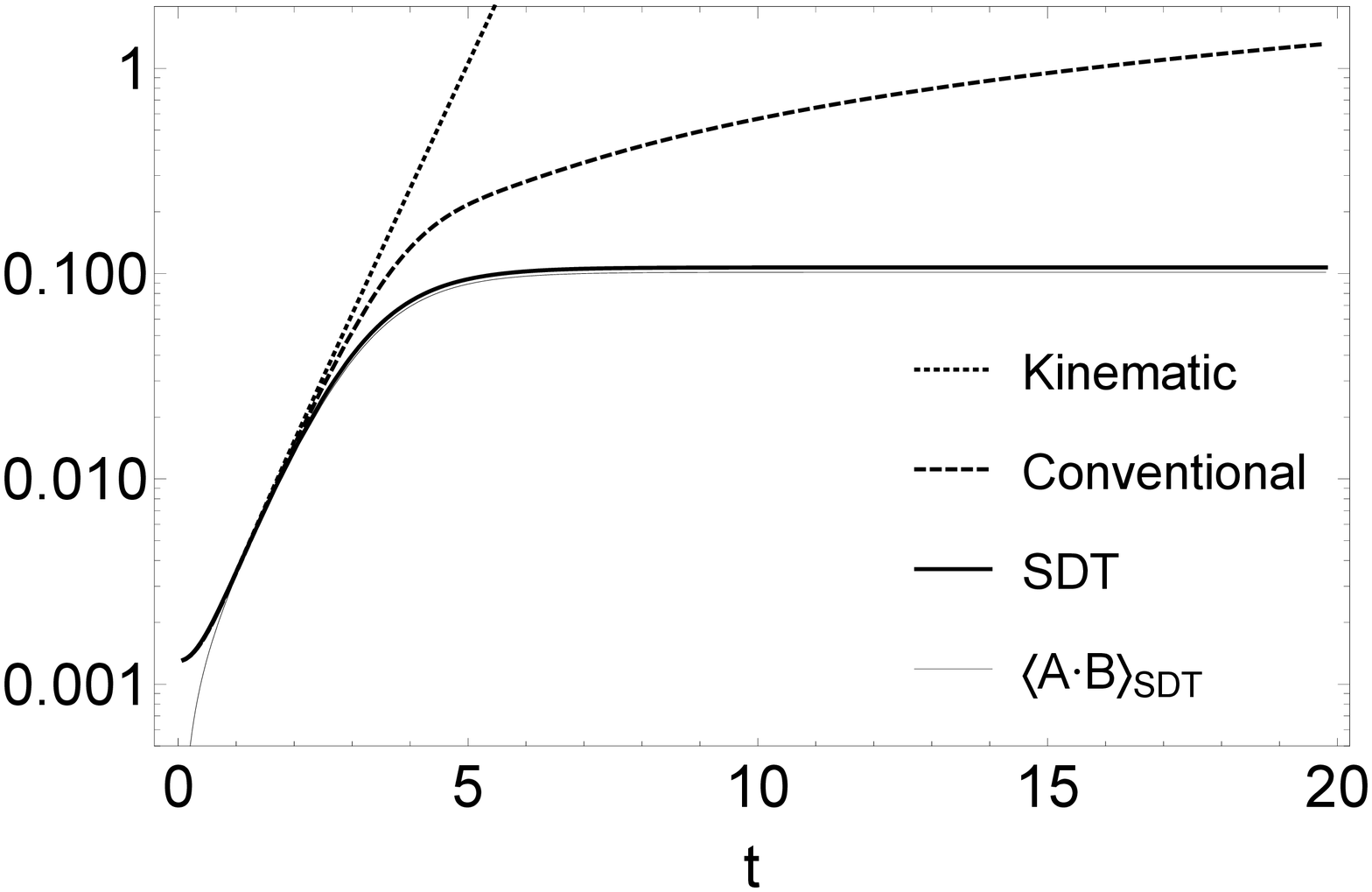}\\
\includegraphics[width=0.8\columnwidth]{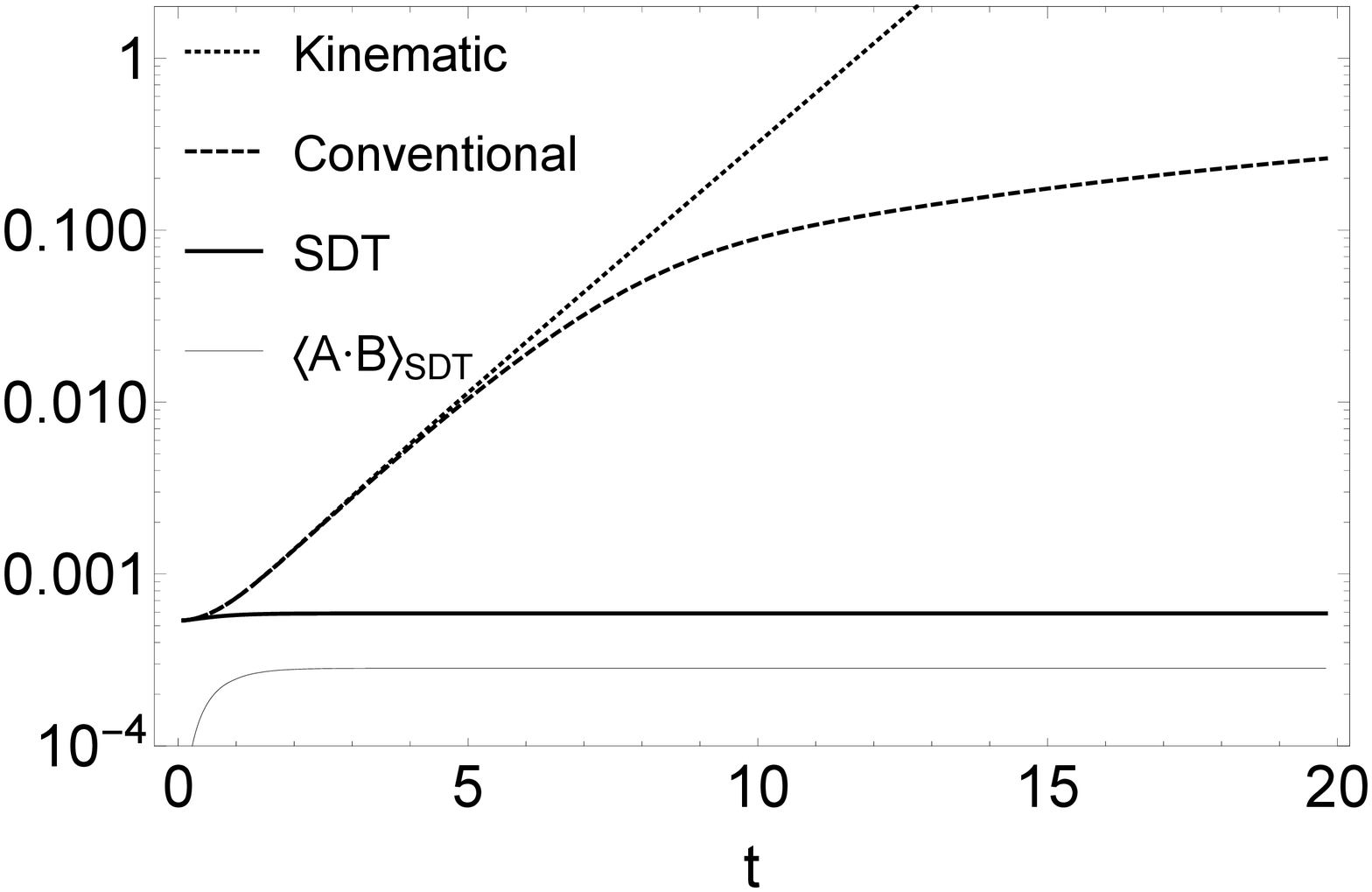}
\caption{
	Time evolution of $\abra{B^2}$ (normalized by $\abra{u^2}$, and $t$ measured in $\sqrt{\abra{u^2}\kL^2}$) without quenching (i.e., kinematic theory, dotted), conventional quenching (dashed), or SDT quenching (solid thick).
	For the SDT case, the time evolution of the spatial-averaged large-scale magnetic helicity $\abra{\bm A\cdot\bmB}$ is also plotted (solid thin).
	Top: helical initial condition (\ref{eqn:initial_alpha_helical}).
	Bottom: non-helical initial condition (\ref{eqn:initial_alpha_nonhelical}).
}
\label{fig:aniso}
\end{figure}

The results are shown in the top panel of figure \ref{fig:aniso}.
While the conventional quenching formula yields an almost linear growth at late time because it only partially offsets the $\bm\alpha$ tensor, our new SDT quenching formalism leads to dynamo saturation.
We have also checked that the total magnetic helicity is conserved up to the precision of the calculation.

Interestingly, at saturation we have $\M{\bm\alpha}=-\K{\bm\alpha}=-\bm\alpha^0$ given equations (\ref{eqn:xi_constraint3}) to (\ref{eqn:zeta}) and that $s\neq0$ in the helical case, which implies that it is still feasible to use helicity conservation to estimate $\abra{B^2}_\text{sat}$ even though the $\bm\alpha$ tensor is anisotropically quenched:
The energy-weighted turbulent wave number is
\beq
k_\text{eff}
=\frac{\int_\kf^\knu dk\ k^{-5/3}}
{\int_\kf^\knu dk\ k^{-8/3}}
\simeq 2\kf,
\label{eqn:keff}
\eeq
and therefore
\beq
\abra{B^2}_\text{sat}\simeq
-\frac{\kL\tr\bm\alpha^0}{\tau_0 (k_\text{eff}/\kf)^{-2/3}k_\text{eff}^2}
\simeq0.1.
\eeq

\subsection{Case of anisotropic traceless $\bm\alpha^0$}
In the third example we use the quenching equations
(\ref{eqn:ddt_alpha_k}) and (\ref{eqn:ddt_alpha_m}) but with the initial condition
$\bm\kL=(1,1,3)/\sqrt{11}$, and
\beq
\K{\tilde{\alpha}}(0)=
\tilde\alpha^0=
\left(
1-
\frac{3k_3^2}{k^2}
\right)
\frac{15\pi^2 \tau F_2(k)}{2k^2},\ 
\M{\tilde{\alpha}}(0)=0
\label{eqn:nonhelical_alpha}
\eeq
which corresponds to
\beq
\bm\alpha^0=\text{diag}\left\{
\frac{1}{2},
\frac{1}{2},
-1
\right\}.
\label{eqn:initial_alpha_nonhelical}
\eeq

The evolution of $\abra{B^2}$ is shown in the lower panel in figure (\ref{fig:aniso}).
There is no significant amplification of the mean magnetic field, consistent with the fact that if $\M{\tilde\alpha}=-\K{\tilde\alpha}=-\tilde\alpha^0$ at saturation, we would have $\bm\alpha|_\text{sat}=\bm0$ and also
\beq
\bar{\bm a\cdot\bmb}|_\text{sat}
=-\abra{\bm A\cdot\bmB}_\text{sat}
=\int\frac{\text{d}^3k}{(2\pi)^3}
\frac{\M{\tilde\alpha_\text{sat}}}{\tau k^2}
=-\int\frac{\text{d}^3k}{(2\pi)^3}
\frac{{\tilde\alpha^0}}{\tau k^2}
=0
\label{eqn:ab_sat}
\eeq
using equation (\ref{eqn:nonhelical_alpha}).
Our result with no large-scale dynamo action does not contradict previous non-helical dynamo studies because they either have focused on a small-scale dynamo \citep{Meneguzzi1981, Haugen2003},
or a low magnetic Reynolds number \citep{Gilbert1988},
or an inhomogeneous flow \citep{Mininni2005}.


\section{Conclusion}
\label{sec:conclusion}
Astrophysical plasmas commonly involve large-scale shear flows and magnetic fields, global rotation, or anisotropic forcing, therefore making turbulent flows anisotropic.
Such anisotropic flows can be either helical or non-helical while maintaining a non-vanishing $\bm\alpha$ tensor in mean-field dynamo models, as long as the flow lacks mirror symmetry.
Interestingly, as we have been shown in section \ref{sec:3}), anisotropic and even non-helical turbulence can grow mean magnetic fields, thereby challenging the breadth of validity of the conventional $\bm\alpha$ quenching formula which is derived for an isotropic helical flow.
{In fact, helicity conservation only employs the isotropic part of the Lorentz back-reaction, and thus provides a fundamentally incomplete theory of quenching for a general anisotropic flow with an $\bm\alpha$ effect.}

We derived a new quenching formalism [equations (\ref{eqn:ddt_alpha_k}) and (\ref{eqn:ddt_alpha_m})] using the selective-damping-$\tau$ (SDT) closure to model correlations with order $\geq 3$.
The new SDT closure is derived from considering how a small-scale dynamo evolves in the absence of the mean field, and subjected to magnetic helicity conservation.
The corresponding physical constraints [equations (\ref{eqn:xi_constraint1}) to (\ref{eqn:xi_constraint3})] are then used as a guide to derive the closure terms.

The SDT quenching conserves magnetic helicity and is applicable to general anisotropic incompressible flows by including the full back-reaction from the Lorentz force and multi-scale nature of turbulence.
We have shown both analytically and numerically that it recovers the conventional quenching in the isotropic case, but in addition, also quenches the field growth in anisotropic flows where the conventional quenching fails to produce complete saturation.
Notably, for isotropic or anisotropic helical cases, estimating the saturated field strength---but not the specific field geometry---can still be estimated by equating small-scale current and kinetic helicities.

We have not taken into account the turbulent diffusion, i.e., the $\bm\beta$ effect in our models.
In the absence of shear, if $\bm\beta$ is included and assumed constant, we would expect it to reduce growth rates, and the dynamo will saturate before $\K{\bm\alpha}=-\M{\bm\alpha}$.
However, our primary conclusion about the $\bm\alpha$ quenching remains unchanged. 
Also note that for the isotropic case $\bm\beta$ is proportional to the turbulent kinetic energy which is approximately a constant with an external forcing, while in general anisotropic flows $\bm\beta$ would have contributions from the turbulent magnetic field and thus incur magnetic back-reaction much like the quenching of $\bm\alpha$.
As such, by assuming $\bm\beta$ to be constant we would only overestimate its influence on dynamo quenching.

A natural extension of the formalism would be to include a large-scale shear flow, the most common source of anisotropy in astrophysical flows.
In the presence of shear, non-helically forced turbulence can also generate helical large-scale magnetic fields, e.g., through a shear-current effect \citep{BrandenburgSubramanian2005} or an inhomogeneous $\bm\alpha$ effect \citep{EbrahimiBlackman2019}.
In these cases a complexity is that at least first-order derivatives of $\bmB$ must be included.
Consequently, in addition to any growth of $\bm \alpha$, quenching of the $\bm\beta$ tensor also has to be considered.
We leave this for future work.

\section*{Acknowledgments}
We thank A. Hubbard for useful comments.
HZ acknowledges support from a Horton Fellowship from the Laboratory for Laser Energetics at UR,
and the 2019 Summer School at the Center for Computational Astrophysics, Flatiron Institute.
The Flatiron Institute is supported by the Simons Foundation.
EB acknowledges support from NSF Grant AST-1813298, KITP UC Santa Barbara funded by NSF Grant PHY-1748958, and Aspen Center for Physics funded by NSF Grant PHY-1607611.

\bibliographystyle{jpp}
\bibliography{sufficientConditionForDynamobib}

\end{document}